\documentstyle{pasj06}
\SetRunningHead{S.Oizumi et al.}{Photometry of V844 Her}

\begin{document}
\title{Long-term monitoring of the short period SU UMa-type dwarf nova,
V844 Herculis}

\author{Shota \textsc{Oizumi}, Toshihiro \textsc{Omodaka}, Hiroyuki
\textsc{Yamamoto}, Shunsuke \textsc{Tanada},\\ Tatsuki \textsc{Yasuda},
Yoshihiro \textsc{Arao}, Kie \textsc{Kodama}, Miho \textsc{Suzuki},
Takeshi \textsc{Matsuo}} 
\affil{Faculty of Science, Kagoshima University,1-21-30 Korimoto,
Kagoshima 890-0065}
\email{oizumi@astro.sci.kagoshima-u.ac.jp}

\author{Hiroyuki \textsc{Maehara}}
\affil{VSOLJ, Namiki 1-13-4,Kawaguchi, Saitama 332-0034}

\author{Kazuhiro \textsc{Nakajima}}
\affil{VSOLJ, 124 Isatotyo, Teradani, Kumano, Mie 519-4673}

\author{Pavol A. \textsc{Dubovsky}}
\affil{Slovak Association of Amateur Astronomers, Podbiel, Slovakia}

\author{Taichi \textsc{Kato}, Akira \textsc{Imada}, Kaori
\textsc{Kubota}, Kei \textsc{Sugiyasu}}
\affil{Department of Astronomy, Faculty of Science, Kyoto University,
Sakyo-ku, Kyoto 606-8502} 

\author{Koichi \textsc{Morikawa}}
\affil{468-3 Satoyamada, Yakage-cho, Oda-gun, Okayama 714-1213}

\author{Ken'ichi \textsc{Torii}}
\affil{Department of Earth and Space Science, Graduate School of
Science, \\ Osaka University, 1-1 Machikaneyama-cho, Toyonaka, Osaka 560-0043}

\author{Makoto \textsc{Uemura}}
\affil{Hiroshima Astrophysical Science Center, Hiroshima University,
Higashi-Hiroshima,\\ Hiroshima 739-8526}

\author{Ryoko \textsc{Ishioka}}
\affil{Subaru Telescope, National Astronomical Observatory of Japan, 650
North A'ohoku Place,\\ Hilo, HI 96720, USA}

\author{Kenji \textsc{Tanabe}}
\affil{Department of Biosphere-Geosphere Systems, Faculty of
Informatics, Okayama University of Science,\\ 1-1 Ridai-cho, Okayama,
Okayama 700-0005 }

\and
\author{Daisaku \textsc{Nogami}}
\affil{Hida Observatory, Kyoto University, Kamitakara, Gifu 506-1314}


\maketitle

\begin{abstract}

 We report on time-resolved CCD photometry of four outbursts of a
 short-period SU UMa-type dwarf nova, V844 Herculis. We successfully
 determined the mean superhump periods to be 0.05584(64) days, and
 0.055883(3) for the 2002 May superoutburst, and the 2006 April-May
 superoutburst, respectively. During the 2002 October observations, we
 confirmed that the outburst is a normal outburst, which is the first
 recorded normal outburst in V844 Her. We also examined superhump period
 changes during 2002 May and 2006 April-May superoutbursts, both of
 which showed increasing superhump period over the course of the plateau
 stage. In order to examine the long-term behavior of V844 Her, we
 analyzed archival data over the past ten years since the discovery of
 this binary. Although photometry is not satisfactory in some
 superoutbursts, we found that V844 Her showed no precursors and
 rebrightenings. Based on the long-term light curve, we further
 confirmed V844 Her has shown almost no normal outbursts despite the
 fact that the supercycle of the system is estimated to be about 300
 days. In order to explain the long-term light curves of V844 Her,
 evaporation in the accretion disk may play a role in the avoidance of
 several normal outbursts, which does not contradict with the relatively
 large X-ray luminosity of V844 Her.

\end{abstract}

\section{Introduction}

Dwarf novae belong to a subclass of cataclysmic variable stars
that consist of a white dwarf (primary) and a late-type star
(secondary). The secondary star fills its Roche lobe and transfers mass
to the primary via inner Lagragian point (L1) and the transferred matter
forms an accretion disk (for a review, see \cite{war95book};
\cite{osa96review}; \cite{hel01book}). Among dwarf novae, there exist
three subtypes based on their light curves. SU UMa-type dwarf novae,
whose orbital period are shorter than 0.1 days in the most cases, are
one of the subtypes, characteristic of exhibiting two types of
outbursts. One is normal outburst, continuing for a few days. The other
is superoutburst, lasting about two weeks, during which modulations
called superhumps are shown. The period of the superhumps are a few
percent longer than that of the orbital period of the system. This is
well explained by a phase-dependent dissipation of a tidally deformed
precessing accretion disk. The most acceptable model for SU UMa stars is
the thermal-tidal instability model developed by \citet{osa89suuma},
well reproducing the majority of observations.

V844 Her was discovered by \citet{ant96newvar} as a variable star near
${\eta}$ Her. \citet{ant96newvar} classified the variable, originally named
Var 43 Her, as a dwarf nova based on the detection of a long
outburst. The light curve in \citet{ant96newvar} is reminiscent of a
superoutburst of SU UMa-type dwarf novae. Time-resolved CCD photometry
was performed by \citet{kat00v844her} during the 1999 September outburst
of V844 Her. Detecting superhumps with a period of 0.05592(2) days, they
firstly confirmed the SU UMa nature of V844 Her. By radial velocity
studies \citet{tho02gwlibv844herdiuma} determined 0.054643(7) days
(78.69 min) as the orbital period of the system. These results indicate
that V844 Her is one of the shortest periods among dwarf novae ever
known. In order to thoroughly investigate the short period system, the
VSNET \citep{kat04vsnet} has placed V844 Her as one of the highest
priorities since the confirmation of the SU UMa nature of the system. In
2002 May, 2002 October, and 2003 October, V844 Her underwent an outburst
and the VSNET Collaboration Team detected superhumps during these
superoutbursts.

On 2006 April 24, Pavol A. Dubovsky reported a brightening of V844 Her
(12.4 mag) to the VSNET ([vsnet-alert 8914]). Thanks to this prompt
report, as well as the seasonal condition of V844 Her (the precise
coordinate is RA: $16^{\rm h} 25^{\rm m} 01^{\rm s}.75$, Dec:
$+39^{\circ} 09' 26''.4$, \cite{2006ApJS..162...38A}), we firstly
succeeded in observing almost the whole superoutburst of V844 Her. The
object is identical with USNO A2.0 1275-8931436 ($B$ = 16.9, $R$ =
16.2). The infrared counterpart of the binary is 2MASS J16250181+3909258
(\cite{hoa02CV2MASS}; \cite{ima06j0137}). V844 Her is also catalogued as
a bright X-ray source by ROSAT, 1RXS J162501.2+390924
\citep{1999A&A...349..389V}.

\section{Observations}

Time-resolved CCD photometry during outburst was performed from 2002 May
20 to 2006 May 20 using 10-100 cm telescopes at 8 sites. The log of
these observations is summarized in table 1. Details of observers are
listed in table 2. In total, we
observed V844 Her for 31 nights, during which 4 outbursts including one
normal outburst were detected. The exposure time was 10-40 seconds. The
read-out time was typically a few seconds. The resultant cadence is much
shorter than the time scale of variations that we focus on. Photometric
data were obtained through the $V$ filter at the Kolonica Saddle
site. The other sites used no filter, which makes the effective
wavelength close to that of $R_{\rm c}$-system. The total data points
amounted to 14674, which is the largest data ever obtained for V844 Her.

After subtracting a dark-current image from the original CCD frames,
flat fielding was performed as the usual manner. The images obtained by
Mhh and KU were processed by the task apphot in IRAF.\footnote{IRAF
(Image Reduction and Analysis Facility) is distributed by National
Optical Observatories, which is operated by the Association of
Universities for Research in Astronomy, Inc., under cooperative
agreement with National Science Foundation.} Kyoto, OUS, RIKEN, and
Okayama data were analyzed by aperture photometry using a Java-based
software developed by one of the authors (TK). Data of Mie were analyzed
using
FitsPhot4.1.\footnote{http://www.geocities.jp/nagai\_kazuo/dload-1.html}
The Maxim DL\footnote{http://www.cyanogen.com/products/maxim\_main.htm}
and the
C$-$MUNIPACK\footnote{http://integral.physics.muni.cz/cmunipack/} were 
used for data obtained at Kolonica Saddle. After correcting systematic
differences between sites, the magnitude was adjusted to that of Kyoto
system except for 2006 observations, for which the magnitude was
adjusted to that of Saitama system. As comparison stars, USNO A2.0
1275-8932542 (RA: $16^{\rm h} 25^{\rm m} 13^{\rm s}.25$, Dec:
$+39^{\circ} 08' 52''.2$, $V$=12.836, $B-V$=0.994) and USNO-A2.0
1292-0262723 (RA: $16^{\rm h} 24^{\rm m} 51^{\rm s}.25$, Dec:
$+39^{\circ} 12' 07''.6$, $V$=12.334, $B-V$=0.662) were used for Kyoto
and Saitama system, respectively
\citep{hen97sequence}.\footnote{ftp://ftp.aavso.org/public/calib/.},
whose constancy was checked by the local stars in the same
images. Heliocentric correction was applied to the observation times
before the following analyses.

\section{Results}

\subsection{2002 May outburst}

\begin{figure}
 \begin{center}
  \resizebox{80mm}{!}{\includegraphics{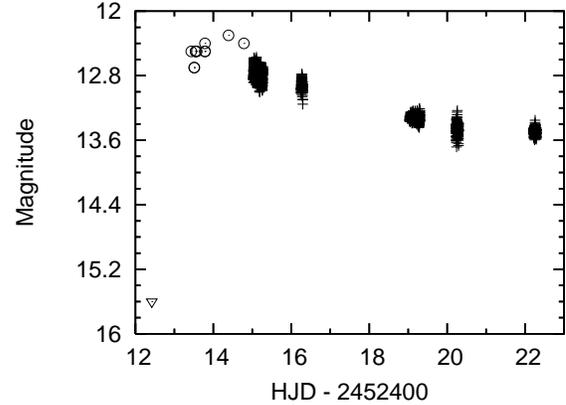}}
 \end{center}
 \caption{The obtained light curve during the 2002 May outburst. The
 abscissa and the ordinate denote the fractional HJD and the magnitude,
 respectively. The bottom triangle means the negative observation. The
 opened circles show the visual observations. The light curve shows no
 sign of precursor.} 
 \label{fig:0205lc}
\end{figure}

The light curve during the 2002 May outburst is presented in figure
\ref{fig:0205lc}. The magnitude declined linearly at the rate of 0.13(1)
mag d$^{-1}$ until HJD 2452420, after which the system kept almost
constant magnitude until the end of our run. Such a halt is sometimes
observed in other SU UMa-type dwarf novae (e.g., V1028 Cyg,
\cite{bab00v1028cyg}. For a comprehensive review, see
\cite{kat03hodel}). On HJD 2452425, 12 days after the detection of the
outburst, V844 Her likely entered the rapid decline stage, when the
visual magnitude was fainter than 15.  

\begin{figure}
 \begin{center}
  \resizebox{80mm}{!}{\includegraphics{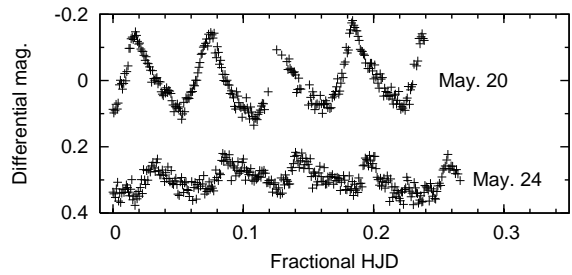}}
 \end{center}
 \caption{Representative light curves obtained during the 2002 May
 superoutburst. The vertical and the horizontal axes denote the
 fractional HJD and differential magnitude, respectively. For the
 purpose of a comparison between nights, the light curve on HJD 2452419
 (May 24) was shifted by 0.3 mag.}
 \label{fig:0205mhp}
\end{figure}

Figure \ref{fig:0205mhp} shows the representative light curves during
the plateau phase after removing daily decline trends for each
run. Rapid rises and slow declines, characteristic of superhumps, are
visible. In order to estimate the superhump period, we applied the phase
dispersion minimization (PDM, \cite{ste78pdm}) method to the prewhitened
light curves during the plateau stage. We determined 0.05584(64)d as
being the best estimated period of the superhump. The error of the
resulting period was estimated using the Lafter$-$Kinman class of
methods as applied by \citet{fer89error}. The obtained superhump period
was in good agreement with the previous studies (\cite{kat00v844her};
\cite{tho02gwlibv844herdiuma}).

\begin{figure}
 \begin{center}
  \resizebox{80mm}{!}{\includegraphics{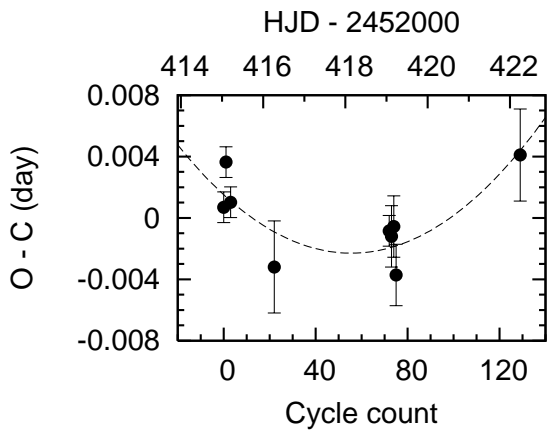}}
 \end{center}
 \caption{$O - C$ diagram of superhump maxima. The $O - C$ was calculated
 against equation (\ref{eqn:02hjdmax}). The dashed curve is the best fitting
 quadratic described in equation (\ref{eqn:02o-c}). Note that the superhump
 period increased as the superoutburst proceeded.}
 \label{fig:0205o-c}
\end{figure}

We extracted the maximum times of superhumps calibrated mainly by
eye. Table \ref{tbl:02o-c} shows the timings of the superhump maxima. A
linear regression to the observed times yields the ephemeris in the
following equation:

\begin{equation}
 \mathrm{HJD(max)} = 2452415.0424(15) + 0.055857(23) \times E,
 \label{eqn:02hjdmax}
\end{equation}

where $E$ is the cycle count of the maximum timings of superhumps, and
the values within the parentheses in the right side of the equation
denote 1-$\sigma$ error, respectively. Using the above equation, we
derived the $O - C$ diagram illustrated in figure \ref{fig:0205o-c}, in
which the dashed line means the best fitting quadratic equation as
follows:

\begin{eqnarray}
 O - C = &1.52(0.99) \times 10^{-3} - 1.37(0.40) \times 10^{-4} \times
  E\nonumber\\\ 
         &+ 1.24(0.34) \times 10^{-6} \times E^{2}.
 \label{eqn:02o-c}
\end{eqnarray}

The quadratic term yields $P_{\rm dot}$ = $\dot{P}$/$P$ =
4.4(1.2)${\times}$10$^{-5}$, indicating that superhump period increases
through the superoutburst. Because of the apparent sparse data listed in
table \ref{tbl:02o-c}, it is likely that the obtained value might
include large uncertainty. Nevertheless, we can conclude the conspicuous
increase in the superhump period during the plateau stage.

\subsection{2002 October outburst}

\begin{figure}
 \begin{center}
  \resizebox{80mm}{!}{\includegraphics{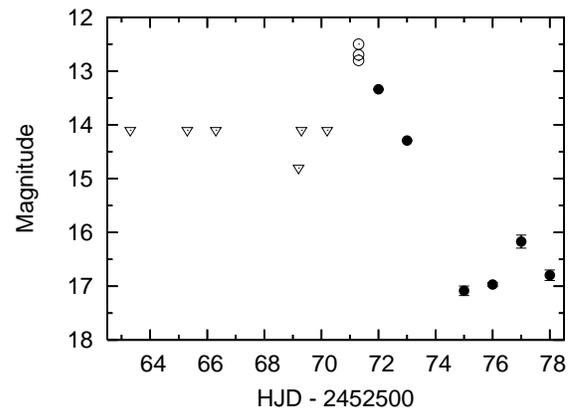}}
 \end{center}
 \caption{Light curve during the 2002 October outburst. The vertical and
 the horizontal axes denote the magnitude and HJD, respectively. The
 filled circles show the nightly averaged magnitudes. The error bars
 mean the standard error for each day. The open circles show visual
 observations. The bottom triangles show the negative observations.}
 \label{fig:0210lc}
\end{figure}

Figure \ref{fig:0210lc} shows the light curve of our run during the 2002
October outburst. The outburst was caught on HJD 2452571, when the
visual magnitude of V844 Her was about 12.5. Our observations started
one night after the detection, when the mean magnitude was about
13.3. Using the data obtained on the first two nights, we determined the
mean decline rate to be 1.20(1) mag d$^{-1}$. The value is large for the
plateau stage of a superoutburst in SU UMa-type dwarf novae. On HJD
2452575, V844 Her faded to 17 mag, which is almost the same as that of
the quiescent magnitude. Based on the negative observation on HJD
2452570, we can estimate that the duration of the outburst was at most
5 days.

\begin{figure}
 \begin{center}
  \resizebox{80mm}{!}{\includegraphics{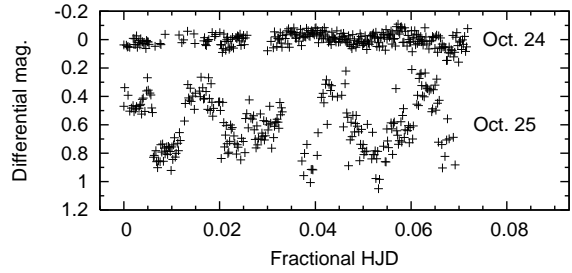}}
 \end{center}
 \caption{Enlarged light curves during the first two nights.
 Hump-like profiles are clearly seen on HJD 2452573 (October 25). The
 profiles of the humps are clearly different from that of superhumps.}
 \label{fig:0210e}
\end{figure}

Figure \ref{fig:0210e} represents the de-trended, enlarged light curves
taken on the first two nights. Interestingly, one can see hump-like
profiles on 2002 October 25 (HJD 2452573) with the amplitude as large as
0.4 mag. They are, however, definitely not superhumps, since three peaks
are detectable with an amplitude of ${\sim}$ 0.4 mag during the 0.06
days run. These results indicate this outburst was the normal
outburst. This is the first recorded $normal$ $outburst$ of V844 Her.

\subsection{2003 October outburst}

\begin{figure}
 \begin{center}
  \resizebox{80mm}{!}{\includegraphics{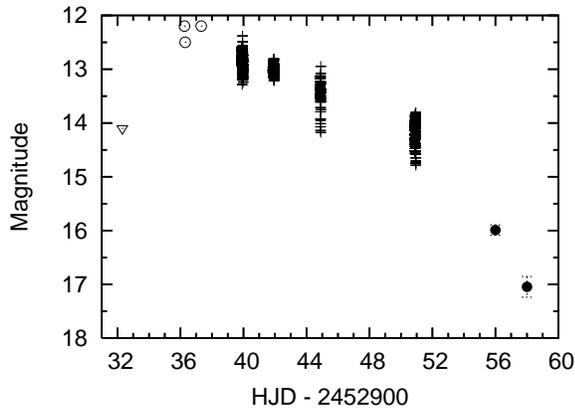}}
 \end{center}
 \caption{The resultant light curve during the 2003 October
 superoutburst. The bottom triangle and the open circles mean the
 negative and the visual observations, respectively.}
 \label{fig:0310lc}
\end{figure}

Figure \ref{fig:0310lc} displays the overall light curves during the
2003 outburst of V844 Her. The duration of the outburst appeared to be
about 2 weeks. The decline rate of 0.12(1) mag d$^{-1}$ until HJD
2452945, which is a typical value for SU UMa-type dwarf novae. Due to
the absence of observations between HJD 2452945 and HJD 2452951, we
cannot specify whether there was a phase of a constant magnitude, as was
observed in the 2002 May superoutburst. After the plateau stage, V844
Her entered the rapid decline phase around HJD 2452956, and the
magnitude returned to its quiescent level on HJD 2452958. No
rebrightenings were observed during our run. Due to the
lack of our observations, we were unable to trace superhump period
change and whether a precursor was present.

\subsection{2006 April outburst}

\subsubsection{light curve}

\begin{figure}
 \begin{center}
  \resizebox{80mm}{!}{\includegraphics{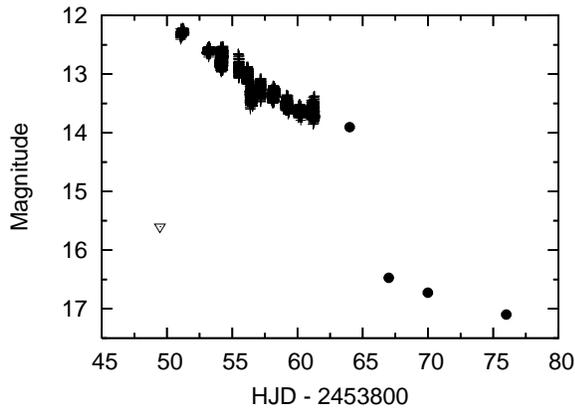}}
 \end{center}
 \caption{The whole light curve of the 2006 April-May superoutburst. The
 horizontal and vertical axes indicate the HJD and the magnitude,
 respectively. The bottom triangle indicates the negative
 observation. The filled circles represent the daily averaged
 magnitude. The typical error bars are within the circles.}
 \label{fig:0604lc}
\end{figure}

The whole light curve of the 2006 April - May outburst is shown in
figure \ref{fig:0604lc}. The duration of the plateau phase was about 2
weeks. The magnitude declined almost constantly at the rate of 0.15(1)
mag d$^{-1}$ from HJD 2453851 to HJD 2453860, after which the magnitude
kept almost constant at the end of the plateau stage. On HJD 2453867,
V844 Her became faint with the magnitude of 16.5. There provided no
evidence of a rebrightening during our run.

\subsubsection{superhump}

\begin{figure}
 \begin{center}
  \resizebox{80mm}{!}{\includegraphics{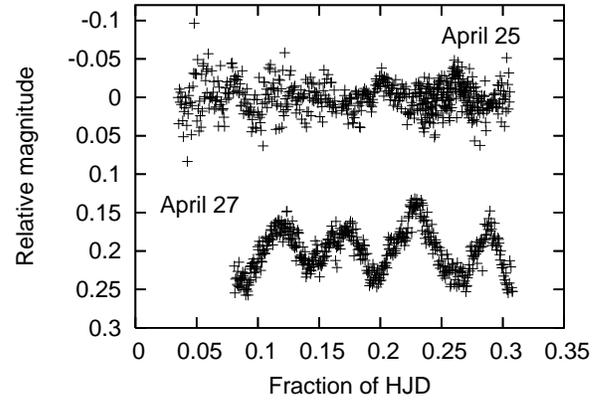}}
 \end{center}
 \caption{First 2-days enlarged light curves since we began observing V844
 Her. There were almost no feature charactaristics on April 25, while hump-like
 profiles appeared clearly on April 27.}
 \label{fig:0604e}
\end{figure}

Figure \ref{fig:0604e} shows the enlarged light curves on the first two
days of our observations. There is no signal of superhumps on HJD
2453851, while there were hump-like modulations on HJD 2453853. However,
their profile suggested that superhumps had not yet fully grown. Then we
suppose the superhumps were detected form HJD 2453854.

We performed a period analysis using the 7473 points between HJD 2453854
and HJD 2453864, after subtracting the linear declining trend. The theta
diagram of the PDM analysis provides the best estimated period of
0.055883(3)d. This value is in good accordance with that obtained during
the 2002 May superoutburst.

\begin{figure}
 \begin{center}
  \resizebox{80mm}{!}{\includegraphics{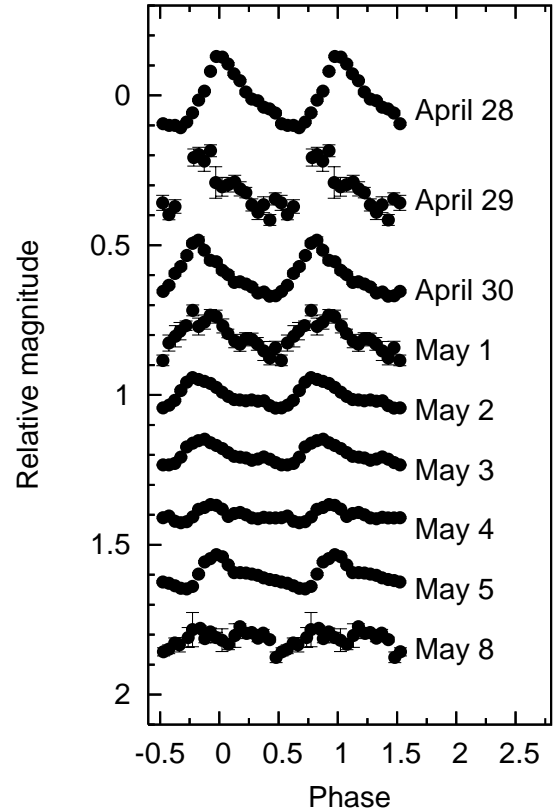}}
 \end{center}
 \caption{Daily averaged light curves of the 2006 April-May superoutburst
 observations.}
 \label{fig:0604dlc}
\end{figure}

Figuer \ref{fig:0604dlc} indicates the daily averaged light curves
during the plateau phase folded by 0.055883(3)d. A rapid rise and slow
decline are a typical feature of superhumps. The data obtained on April
28(HJD 2453854) showed the superhumps with an amplitude of 0.2 mag, then
the superhump amplitude decreased gradually. A hint of the regrowth of
the superhump can be seen on May 5(HJD 2453861).

\subsubsection{superhump period change}
The superhump maximum timings measured by eye are listed in table 4. A
linear regression yields the following equation on the superhump maximum
timings;

\begin{equation}
 \mathrm{HJD(max)} = 2453854.1284(14) + 0.055885(18) \times E.
 \label{eqn:06hjdmax}
\end{equation}

\begin{figure}
 \begin{center}
  \resizebox{80mm}{!}{\includegraphics{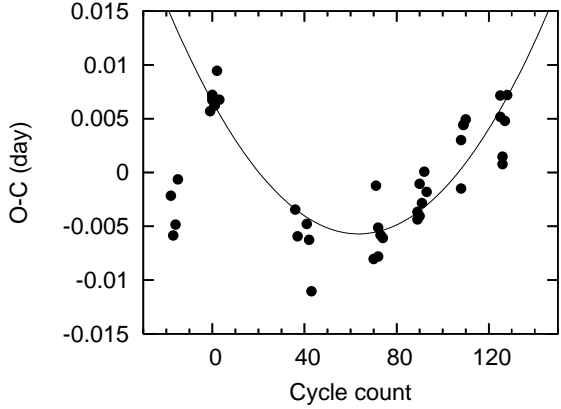}}
 \end{center}
 \caption{$O-C$ diagram during the 2006 superoutburst. The horizontal and
 vertical axes denote the cycle count (E) and $O-C$ (day),
 respectively. The solid curve is the best fitting quadratic described
 in equation (\ref{eqn:06o-c}). Note that the superhump period increased
 as the superoutburst proceeded.}
 \label{fig:0604o-c}
\end{figure}

The obtained $O - C$ diagram is exhibited in figure
\ref{fig:0604o-c}. For $-$1 \textless $E$ \textless 128, the best fitting
quadratic equation is given by:

\begin{eqnarray}
 O - C = &6.6(0.9) \times 10^{-3} - 3.88(0.34) \times 10^{-4} \times
  E\nonumber\\ 
         &+ 3.05(0.27) \times 10^{-6} \times E^{2}.
 \label{eqn:06o-c}
\end{eqnarray}

This equation yields $P_{\rm dot}$ = $\dot{P}$/$P$=10.9(1.0)
${\times}$10$^{-5}$, meaning that the superhump period increases
through the superoutburst.

\subsection{distance and X-ray luminosity}

It is well known that an accurate estimation of the distance to the
dwarf novae is not easy. Nevertheless, we can roughly estimate it
using an empirical relation derived by \citet{war87CVabsmag} as follows:

\begin{equation}
 M_{\rm V} = 5.64 - 0.259P,
 \label{eqn:Mv}
\end{equation}

where $M_{\rm V}$ is the absolute magnitude at the maximum during a
normal outburst, and $P$ is the orbital period of the system in the unit
of hours. The above equation could be applied for the systems which have
a low inclination and do not reach the period minimum. Based on the
previous investigations by \citet{tho02gwlibv844herdiuma}, both
conditions can be satisfied for V844 Her. Substituting $P$=1.3115 into
equation (\ref{eqn:Mv}) and with a little algebra (here we assume that
the maximum $V$ magnitude of V844 Her is 12.6), we roughly derived
d=290(30) pc as an estimated distance.

Using the obtained distance, we can also estimate the X-ray luminosity
of V844 Her following the same manner as \citet{ver97ROSAT}, who showed
that the $ROSAT$ PSPC countrate in channel 52-201 corresponds to a flux in
the 0.5-2.5 keV bandpass given by

\begin{equation}
 {\rm log}F_{\rm 0.5-2.5keV}({\rm erg cm^{-2}s^{-1}}) \sim {\rm log
 cr_{52-201}(s^{-1})} - 10.88,
 \label{eqn:x-ray}
\end{equation}

where cr$_{\rm 52-201}$ denotes the countrate in channel 52-201. With a
little algebra, we can obtain the X-ray luminosity between 0.5-2.5 keV
is 10$^{31.0{\pm}0.2}$ erg s$^{-1}$. Although the observed X-ray
luminosity will be affected by some effects including the inclination of
the system, the derived value is relatively large compared to other SU
UMa-type dwarf novae given by \citet{ver97ROSAT}. 

\section{Discussion}

\subsection{superhump period change}

Historically, the superhump period had been known to
decrease during the course of the superoutburst before the tidal
instability was discovered (\cite{1979A&A....77....7H}; \cite{vog83lateSH}).
The decrease of superhump period was ascribed to shrinkage of the disk
radius, or simply a natural consequence of mass depletion from the
accretion disk \citep{osa85SHexcess}. Recently, particularly over the
past decade, the picture has been altered since numerous systems showed
an increase of the superhump period. Such systems are mainly WZ Sge-type
dwarf novae, and SU UMa-type dwarf novae with short orbital periods
(\cite{sem97swuma}; \cite{nog98swuma}; \cite{bab00v1028cyg};
\cite{ish01rzleo}; \cite{uem02j2329}; \cite{ole03v1141aql};
\cite{nog04vwcrb}; \cite{ima05gocom};
\cite{tem06asas0025}).\footnote{An increase of the superhump period was
originally discovered in OY Car \citep{krz85oycarsuper}, which had been
left behind for a long time.} Observationally, there appears to be a
borderline around $P_{\rm sh}$=0.063 days below which the superhump period
tends to increase \citep{ima05gocom}.

\begin{figure}								
 \begin{center}								
  \resizebox{80mm}{!}{\includegraphics{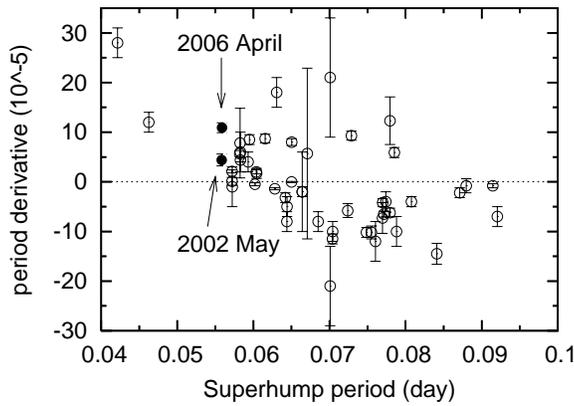}}			
 \end{center}								
 \caption{The $\dot{P}$/P diagram for SU UMa-type dwarf nova. The data
 points were reference to \citet{uem05tvcrv}. The filled circles denote
 the data obtained in this work.}
 \label{fig:p-d}
\end{figure}

Figure \ref{fig:p-d} illustrates the superhump period derivative against
the mean superhump period of SU UMa-type dwarf novae. The value for V844
Her is pointed with the filled circles. In figure \ref{fig:p-d}, V844
Her is likely to lie in the general trend. Hence, we firstly confirmed
that V844 Her showed the positive $P_{\rm dot}$ derivative and became
the shortest period SU UMa-type dwarf nova that was confirmed to exhibit
an increase of the superhump period.\footnote{Here we exclude two
systems, V485 Cen and EI Psc, for which the systems are believed to pass
through another evolutional sequence \cite{uem02j2329}; \cite{pod03amcvn}.}

Additionally, we should briefly note on figure \ref{fig:0604o-c}, where
one can see data points departured from the quadratic equation
(\ref{eqn:06o-c}), corresponding to $E$ ${\sim}$ $-$20. Recent
CCD photometry indicates that this feature is observed not only for V844
Her, but also for other short period SU UMa-type stars exhibiting the
positive $P_{\rm dot}$. The systems include ASAS 102522-1542.4 (Maehara
et al. in preparation), FL TrA (Imada et al. (2006)), ASAS 160048-4846.2
(\cite{ima06asas1600letter}; Imada et al. in preparation). Theoretical
models suggest a dramatic variation in temperature or pressure in the
accretion disk is a possible cause of the superhump period change in
this stage (\cite{mur98SH}; \cite{mon01SH};
\cite{2006MNRAS.371..235P}). We require further samples in order to
discuss the nature of the superhump period change at this early stage.

\subsection{on the nature of V844 Her}

It is well known that SU UMa-type dwarf novae show two types of
superoutbursts: superoutbursts with a precursor, and superoutbursts
without a precursor, though precursor-main superoutburst is hardly
observed. For superoutbursts without a precursor, a handful of systems,
especially archetype TOADs WX Cet and SW UMa, show three types of
superoutburst: a short superoutburst with a duration as short as 10
days, an intermediate superoutburst continuing for 2 weeks, and a long
superoutburst lasting longer than 20 days \citep{how95swumabcumatvcrv}.
In the case of V844 Her, all the three superoutbursts reported here
lasted about 2 weeks, suggesting that these superoutbursts belong to the
intermediate category, or V844 Her simply lies in the majority of SU
UMa-type dwarf novae.

In order to examine whether V844 Her shows other types of
superoutbursts, we extracted the observations reported to AAVSO and
VSNET since the 1996 discovery. Table \ref{tbl:pre_outbursts} summarizes
the recorded outbursts of V844 Her, from which we can properly give a
constraint on the durations of outbursts\footnote{As for data obtained
before 1996, \citet{ant96newvar}
studied Moscow collection of photographic plates and found four
outbursts of V844 Her. Two were definitely superoutburst, of which one
was superoutburst without a precursor.}. As can be noticed in table
\ref{tbl:pre_outbursts}, no superoutburst provides evidence for a
duration longer than 20 days, and the durations of the superoutbursts
appears to converge to 2 weeks. From the archives, we newly
discovered two facts. One is that V844 Her shows no
precursor. Of course we have overlooked the onset of superoutburst in a
few cases, for which we cannot specify the type of
superoutburst. However, the absence of a precursor has been confirmed in
the most cases of the superoutbursts by virtue of the amateur
observers. The other is that no rebrightenings have been observed in
V844 Her despite careful monitoring of the system since the 1996 discovery
\citep{ant96newvar}. Recent extensive studies over the past decades
suggest that rebrightenings tend to occur among SU UMa-type dwarf novae
with short orbital periods (\cite{kuu96TOAD}; \cite{ima06j0222}).

The most interesting fact is, although it has been mentioned for a long
time, that V844 Her shows almost no normal outbursts (\cite{kat00v844her};
\cite{tho02gwlibv844herdiuma}). From the viewpoint of the original
thermal-tidal instability model, normal outbursts occur more frequently
as the mass transfer rate from the secondary increases
(\cite{osa89suuma}; \cite{osa95eruma}). Further, the optical spectrum of
V844 Her suggests a relatively high mass transfer rate
\citep{2005AJ....129.2386S}, which accelerates a circulation on the limit
cycle in the ${\Sigma}$-$T$ diagram. From table \ref{tbl:pre_outbursts},
the supercycle of V844 Her is estimated to be about 300 days, which is
in agreement with the previous work \citep{kat00v844her}. When compared
to the other systems having similar supercycles, e.g., Z Cha, the
peculiarity of V844 Her becomes much clear with respect to the absence
of normal outbursts.

Although we cannot draw a firm conclusion against the infrequent normal
outbursts, one possibility is that evaporation in the accretion disk is
working well so that a hole is created in the inner region of the disk and
avoids an outburst (\cite{mey94siphonflow}; \cite{liu95DNevaporation};
\cite{las95wzsge}; \cite{ham97wzsgemodel}; \cite{min98wzsge}). The model
also suggests large X-ray luminosity \citep{las95wzsge} and expansion of
the accretion disk during quiescence \citep{min98wzsge}. As for the
former, the relatively large X-ray luminosity of V844 Her may be
suggestive of the evaporation. The latter should be investigated by
peak-separation studies during quiescence.

\section{Conclusions}

In this paper, we reported on time resolved CCD photometry during 2002
May, 2002 October, 2003 October, and 2006 April outbursts, of which
three were superoutburst. We estimated the mean superhump periods of
0.05584(64)d for 2002 May and 0.055883(3)d for 2006 April
superoutbursts, respectively. We successfully
examined superhump period change during the 2002 May and 2006 April
superoutburst. The resultant period derivatives showed an period increase
of superhumps during the plateau stages in both superoutbursts, which we
confirmed in V844 Her for the first time. We also derived the distance
to V844 Her to be 290(30) pc. Using the value, the X-ray luminosity
between 0.5 and 2.5 keV is estimated 10$^{31.0{\pm}0.2}$ erg s$^{-1}$.

To appreciate the long-term behavior of V844 Her, we investigated the
archival light curves since the discovery of the variable. Using the
extensive data, we estimated a possible supercycle to be ${\sim}$ 300
days, which is in good agreement with
\citet{tho02gwlibv844herdiuma}. From the archives, it turned out that
V844 Her shows neither precursors nor rebrightenings, although we cannot
rule out the possibility that we missed a few events. We also confirmed
that V844 Her shows almost no normal outbursts in spite of the
intermediate supercycle among SU UMa-type dwarf novae. A possible
explanation on the absence of a normal outburst is that the evaporation
mechanism may play a role, which is consistent with the relatively large
X-ray luminosity of V844 Her. In the future, the evolution of the disk
radius during quiescence should be investigated in order to test our
suggestion.

\vskip 10mm

We are grateful to many observers who have reported vital observations.
We acknowledge with thanks the variable star observations from the AAVSO
and VSNET International Database contributed by observers worldwide and
used in this research. This work is supported by a Grant-in-Aid for the
21st Century COE ``Center for Diversity and Universality in Physics''
from the Ministry of Education, Culture, Sports, Science and Technology
(MEXT). This work is partly supported by a grant-in aid from the
Ministry of Education, Culture, Sports, Science and Technology
(No. 17340055, 16340057, 17740105). Part of this work is supported by a
Research Fellowship of the Japan Society for the Promotion of Science
for Young Scientists (RI).

\begin{table*}
 \caption{Log of observations}
 \label{tbl:obs}
 \begin{center}
  \begin{tabular}{rccccc}
 \hline\hline
 Date & $^*$ HJD start  & $^*$ HJD end & $^\dagger$Exp & $^\ddagger$ N & $^\S$ ID \\ 
\hline
2002 May      20 & 52415.0264 & 52415.2657 & 10 & 1051 & KM \\
              21 & 52416.2469 & 52416.2942 & 30 & 140 & Kyoto \\
              24 & 52419.0356 & 52419.3031 & 10 & 1244 & KM \\
                 & 52419.2251 & 52419.2973 & 30 & 182 & Kyoto \\
              25 & 52420.2194 & 52420.2997 & 30 & 215 & Kyoto \\
              27 & 52422.2133 & 52422.2934 & 30 & 195 & Kyoto \\
2002 October  24 & 52571.8704 & 52571.9989 & 30 & 122 & Kyoto \\
                 & 52571.9074 & 52571.9792 & 10 & 290 & OUS \\
              25 & 52572.8721 & 52573.0185 & 30 & 261 & Kyoto \\
                 & 52572.9032 & 52572.9724 & 10 & 257 & OUS \\
              27 & 52574.9060 & 52574.9557 & 40 & 89  & RIKEN \\
              28 & 52575.8676 & 52575.9966 & 40 & 222 & RIKEN \\
                 & 52575.8758 & 52575.9884 & 30 & 188 & Kyoto \\
              29 & 52576.8715 & 52576.9917 & 30 & 122 & Kyoto \\
                 & 52576.9266 & 52576.9620 & 25-30 & 77 & OUS \\ 
              30 & 52577.8781 & 52577.9332 & 30 &  91 & Kyoto \\
2003 October  27 & 52939.8714 & 52939.9845 & 30 & 185 & Kyoto \\
              29 & 52941.8702 & 52941.9478 & 30 & 148 & Kyoto \\ 
2003 November  1 & 52944.8663 & 52944.9188 & 30 &  65 & Kyoto \\
               7 & 52950.8707 & 52950.9481 & 30 & 116 & Kyoto \\
              12 & 52955.8781 & 52955.9428 & 30 &  93 & Kyoto \\
              14 & 52957.8631 & 52957.8721 & 30 &  18 & Kyoto \\
2006 April    25 & 53851.0347 & 53851.1642 & 30 & 235 & Njh \\
                 & 53851.1684 & 53851.2948 & 30 & 278 & Kyoto \\
                 & 53851.2312 & 53851.3075 & 30 & 124 & Mhh \\
              27 & 53853.0808 & 53853.3080 & 30 & 533 & Kyoto \\
              28 & 53854.0403 & 53854.1606 & 30 & 221 & Njh \\
                 & 53854.1103 & 53854.2534 & 30 & 408 & Mhh \\
                 & 53854.1183 & 53854.3170 & 30 & 373 & Kyoto \\
              29 & 53855.4640 & 53855.5378 & 30 & 107 & PD \\
              30 & 53856.1070 & 53856.2445 & 30 & 506 & Mhh \\
                 & 53856.3505 & 53856.5867 & 30 & 343 & PD \\
2006 May       1 & 53857.1459 & 53857.1463 & 30 & 200 & Mhh \\
               2 & 53858.0236 & 53858.1938 & 30 & 307 & Njh \\
                 & 53858.1229 & 53858.3097 & 30 & 383 & Kyoto \\
               3 & 53859.0606 & 53859.1608 & 30 & 182 & Njh \\
                 & 53859.0655 & 53859.2031 & 30 & 363 & Mhh \\
                 & 53859.1268 & 53859.2947 & 30 & 192 & Kyoto \\
                 & 53859.1568 & 53859.3309 & 30 & 439 & KU \\
               4 & 53860.0693 & 53860.2805 & 30 & 534 & Mhh \\
                 & 53860.0842 & 53860.2142 & 30 & 237 & Njh \\
                 & 53860.1026 & 53860.3091 & 30 & 498 & Kyoto \\
                 & 53860.1559 & 53860.3223 & 30 & 369 & KU \\
               5 & 53861.0016 & 53861.2309 & 30 & 406 & Njh \\
                 & 53861.0656 & 53861.3024 & 30 & 538 & Kyoto \\
                 & 53861.0755 & 53861.3121 & 30 & 452 & KU \\
                 & 53861.1565 & 53861.2755 & 30 & 312 & Mhh \\
               8 & 53864.1508 & 53864.2961 & 30 & 184 & Kyoto \\
              11 & 53867.0282 & 53867.2747 & 30 & 469 & Kyoto \\
              14 & 53870.0300 & 53870.1547 & 30 & 230 & Kyoto \\
              20 & 53876.1544 & 53876.3008 & 30 & 243 & Kyoto \\
 \hline
 \multicolumn{6}{l}{$^*$ : Start and end time of the observation. HJD $-$
 2400000.} \\
 \multicolumn{6}{l}{$^\dagger$ : Exposure time} \\
 \multicolumn{6}{l}{$^\ddagger$ : Number of frames.} \\
 \multicolumn{6}{l}{$^\S$ : ID of the observers. See Table
 \ref{tbl:observers}.} \\
  \end{tabular}
 \end{center}
\end{table*}

\begin{table*}
 \caption{List of observers.}
 \label{tbl:observers}
 \begin{center}
  \begin{tabular}{cccc}
   \hline\hline
   ID & Observer & Site & Telescopes \\
   \hline
   KM & K. Morikawa & Okayama, Japan & 25cm \\
   KU & S. Oizumi $+$ $^*$ & Kagoshima, Japan & 100cm \\
   Kyoto & A. Imada $+$ $^\dagger$ & Kyoto, Japan & 40cm \\
   Mhh & H. Maehara & Saitama, Japan & 25cm \\
   Njh & K. Nakajima & Mie, Japan & 25cm \\
   OUS & K. Tanabe & Okayama, Japan & 10cm \\
   PD & P. A. Dubovsky & Kolonica Saddle, Slovakia & 28cm $\&$ 30cm \\
   RIKEN & K. Torii & Saitama, Japan & 25cm \\
   \hline
   \multicolumn{4}{l}{$^*$ : observer S.Oizumi, H.Yamamoto, S.Tanada, T.Yasuda,
   Y.Arao, K.Kodama, M.Suzuki and T.Matsuo.} \\
   \multicolumn{4}{l}{$^\dagger$ : observer A. Imada, K. Kubota,
   K. Sugiyasu and T. Kato.} \\
  \end{tabular}
 \end{center}
\end{table*}

\begin{table}
 \caption{Timings of superhump maxima during the 2002 May superoutburst.}
 \label{tbl:02o-c}
 \begin{center}
 \begin{tabular}{ccccc}
 \hline\hline
 E$^*$ & HJD$^{\dagger}$ & $O-C$ & Error$^{\ddagger}$ & ID \\
 \hline
  0 & 2415.0431 &  0.000700 & 0.001 & KM \\
  1 & 2415.1019 &  0.003641 & 0.001 & KM \\
  3 & 2415.2110 &  0.001023 & 0.001 & KM \\
 22 & 2416.2681 & -0.003195 & 0.003 & Kyoto \\
 72 & 2419.0634 & -0.000839 & 0.001 & KM \\
 73 & 2419.1189 & -0.001198 & 0.002 & KM \\
 74 & 2419.1754 & -0.000557 & 0.002 & KM \\
 75 & 2419.2281 & -0.003716 & 0.002 & KM \\
 129 & 2422.2523 &  0.004105 & 0.003 & Kyoto \\
 \hline
\multicolumn{3}{l}{$^*$ Cycle count.} \\
\multicolumn{3}{l}{$^{\dagger}$ HJD-2450000} \\
\multicolumn{3}{l}{$^{\ddagger}$ In the unit of day.} \\
 \end{tabular}
 \end{center}
\end{table}

\begin{longtable}{ccccc}
 \caption{Timings of the superhump maxima during the 2006 April
 superoutburst.}
 \label{tbl:06o-c}\\
 \hline\hline
 E$^*$ & HJD$^{\dagger}$ & $O-C$ & Error$^{\ddagger}$ & ID \\ 
 \hline
 -18 & 3853.1201 & -0.002163 & 0.006 & Kyoto \\
 -17 & 3853.1723 & -0.005855 & 0.007 & Kyoto \\
 -16 & 3853.2292 & -0.004846 & 0.005 & Kyoto \\
 -15 & 3853.2893 & -0.000638 & 0.005 & Kyoto \\
  -1 & 3854.0782 &  0.005781 & 0.004 & Njh \\
   0 & 3854.1351 &  0.006789 & 0.001 & Mhh \\
     & 3854.1354 &  0.007089 & 0.002 & Njh \\
     & 3854.1356 &  0.007289 & 0.002 & Kyoto \\
   1 & 3854.1904 &  0.006298 & 0.002 & Kyoto \\
     & 3854.1905 &  0.006298 & 0.002 & Mhh \\
   2 & 3854.2495 &  0.009506 & 0.005 & Kyoto \\
   3 & 3854.3027 &  0.006815 & 0.003 & Kyoto \\
  36 & 3856.1368 & -0.003606 & 0.002 & Mhh \\
  37 & 3856.1901 & -0.006098 & 0.002 & Mhh \\
  41 & 3856.4149 & -0.004964 & 0.003 & PD \\
  42 & 3856.4693 & -0.006456 & 0.009 & PD \\
  43 & 3856.5204 & -0.011247 & 0.006 & PD \\
  70 & 3858.0322 & -0.008419 & 0.019 & Njh \\
  71 & 3858.0950 & -0.001610 & 0.009 & Njh \\
  72 & 3858.1442 & -0.008202 & 0.004 & Njh \\
     & 3858.1470 & -0.005502 & 0.004 & Kyoto \\
  73 & 3858.2022 & -0.006193 & 0.004 & Kyoto \\
  74 & 3858.2577 & -0.006485 & 0.005 & Kyoto \\
  89 & 3859.0978 & -0.004858 & 0.003 & Mhh \\
     & 3859.0984 & -0.004158 & 0.005 & Njh \\
  90 & 3859.1539 & -0.004550 & 0.003 & Mhh \\
     & 3859.1570 & -0.001550 & 0.005 & Kyoto \\
  91 & 3859.2111 & -0.002837 & 0.003 & KU \\
  92 & 3859.2699 & -0.000078 & 0.002 & KU \\
  93 & 3859.3239 &  0.001807 & 0.005 & KU \\
 108 & 3860.1624 & -0.002097 & 0.008 & Kyoto \\
     & 3860.1670 &  0.003014 & 0.002 & KU \\
 109 & 3860.2242 &  0.003811 & 0.010 & Kyoto \\
     & 3860.2243 &  0.004429 & 0.002 & KU \\
 110 & 3860.2806 &  0.004319 & 0.005 & Kyoto \\
 125 & 3861.1191 &  0.004446 & 0.007 & Kyoto \\
     & 3861.1212 &  0.007165 & 0.003 & KU \\
 126 & 3861.1706 &  0.000055 & 0.012 & Kyoto \\
     & 3861.1714 &  0.000755 & 0.003 & Mhh \\
 127 & 3861.2305 &  0.004063 & 0.004 & Mhh \\
 128 & 3861.2889 &  0.007209 & 0.003 & KU \\
 \hline
\multicolumn{3}{l}{$^*$ Cycle count.} \\
\multicolumn{3}{l}{$^{\dagger}$ HJD-2450000} \\
\multicolumn{3}{l}{$^{\ddagger}$ In the unit of day.} \\
\end{longtable}

\begin{table}
 \caption{Recorded outbursts of V844 Her.}
 \label{tbl:pre_outbursts}
 \begin{center}
  \begin{tabular}{lccccc}
   \hline\hline
   Date & & Duration & Type & precursor\\ 
   \hline
   1996 & October    & 7$<T<$20 & S & $\times$  \\
   1997 & May        &  $T=$15 & S & $\times$ \\
   1998 & December   & 8$<T<$19 & S & ? \\
   1999 & September  & 14$<T<$16 & S & $\bigtriangleup$ \\
   2000 & July       & 16$<T<$17 & S & $\times$ \\
   2001 & August     & 16$<T<$18 & S & $\times$ \\
   2002 & May        & 12$<T<$15 & S & $\times$ \\
   2002 & October    &  5$<T<$6 & N & -- \\
   2002 & December   & 10$<T<$18  & S & ? \\
   2003 & May        &  3$<T<$5  & N & -- \\
   2003 & October    & 12$<T<$16 & S & ? \\
   2005 & January    & 14$<T<$16 & S & $\bigtriangleup$ \\
   2006 & April      & 15$<T<$18 & S & $\times$ \\
   \hline
   \multicolumn{5}{l}{S : Superoutburst.} \\
   \multicolumn{5}{l}{N : Normal outburst.} \\
   \multicolumn{5}{l}{$\times$ : No precursor.} \\
   \multicolumn{5}{l}{$\bigtriangleup$ : Probably no precursor.} \\
   \multicolumn{5}{l}{? : Unable to discern the type of superoutburst.} \\
  \end{tabular}
 \end{center}
\end{table}

\end{document}